# Genetically programmable optical random neural networks

Bora Çarpınlıoğlu, Uğur Teğin*

Department of Electrical and Electronics Engineering, Koç University, Istanbul, 34450, Türkiye
*utegin@ku.edu.tr

**Abstract**
Today, machine learning tools, particularly artificial neural networks, have become crucial for diverse applications. However, current digital computing tools to train and deploy artificial neural networks often struggle with massive data sizes and high power consumptions. Optical computing provides inherent parallelism accommodating high-resolution input data and performs fundamental operations with passive optical components. However, most of the optical computing platforms suffer from relatively low accuracies for machine learning tasks due to fixed connections while avoiding complex and sensitive techniques. Here, we demonstrate a genetically programmable yet simple optical neural network to achieve high performances with optical random projection. By genetically programming the orientation of the scattering medium which acts as a random projection kernel and only using 1% of the search space, our novel technique finds an optimum kernel and improves initial test accuracies by 8-41% for various machine learning tasks. Through numerical simulations and experiments on a number of datasets, we validate the programmability and high-resolution sample processing capabilities of our design. Our optical computing method presents a promising approach to achieve high performance in optical neural networks with a simple and scalable design.

## Introduction

Nowadays, artificial neural networks are heavily used for learning from data to perform intricate tasks of high complexity(1). In the last decade, machine learning applications have started to cover diverse fields, from autonomous driving to healthcare diagnosis, and demonstrate outstanding success in real-life applications(2). With the availability of high volumes of data, the trend of machine learning is to develop computationally complex and data-hungry giant models. This increasing complexity in machine learning models adds computational and environmental challenges during training and inference steps(3). Moreover, the currently used digital computing platforms, which are based on Turing–von Neumann architecture, are notoriously limited by the memory access bottleneck and cannot keep up with the demands of large models efficiently. The current transistor-based computing methods are reaching the physical and technological limits, indicating the end of Moore's Law.

Since all the information in the macroscopic world is natively analog, analog computing is an intuitive alternative hardware solution for creating and deploying machine learning models. Unlike digital computing, which processes information in discrete values (0s and 1s), analog computing operates on continuous signals; thus, it is compatible with complex, high-dimensional data. However, analog computing is intrinsically noisy and not as precise as digital computing. The neural networks' relative robustness against noise and device imperfections makes them ideal applications for analog computing solutions. In recent decades, analog computing tools such as brain-inspired (neuromorphic) computing methods, including approaches like reservoir computing(4) and extreme learning machines(5,6), emerged. With fixed connections in the reservoir or hidden layers, these methods provide decent accuracy with less training effort.

Optical computing, a branch of analog computing, serves as a prominent candidate to accelerate machine learning efforts by offering several key operations like Fourier transform, convolution operations, and matrix multiplications intrinsically(7). This rich framework has led to the development of optical computers capable of pattern recognition(8) and the optical implementation of neural network models(9) rapidly. However, initial efforts faced challenges in controlling/training optical connections and implementing nonlinear connections. With the introduction of neuromorphic computing methods in the 2000s, the training problem of optical computing has been circumvented. Recently, interest in optical neural networks has grown with free space, fiber, and integrated designs(10–13). In its simple form, implementing artificial neural networks efficiently and rapidly in optics can be achieved by utilizing optical diffraction, metasurfaces, wave propagation, and passive optical elements like lenses and/or scattering media(14,15). Especially, using a disordered medium comprising millions of fully connected photonic nodes serves as a static random reservoir(16) with random matrix multiplication and offers random mapping of data for dimensionality reduction(17). By incorporating digital feedback loops to the fixed random mapping of a disordered medium, optical neural networks are demonstrated for chaotic system prediction(18,19).

In order to control light propagation in disordered media, the transmission matrix (TM) formalism is sometimes adopted. However, due to the unavailability of phase information, a high number of modes, and all the angles and polarizations, such a characterization is difficult in practice(17,20), and computational approaches have been developed(21,22) to ease the process of manipulating the TM.

Here, we present the first genetically programmable optical random neural network without the computational overload of determining the full TM and improve the performance of scattering-based optical computing platforms for machine learning applications. Our novel method is based on finding an optimum random projection kernel to map information optically for designated machine learning applications by altering the disordered medium (diffuser) orientation inside the optical neural network.

We employ an evolutionary search algorithm (genetic algorithm) and significantly increase accuracy for various datasets by using only 1% of the search space to find optimum random projection kernels. Our simple programmable optical neural network has been tested in numerical simulations and experiments using Breast MNIST(23), Covid-19 X-Ray(24), and retinal disease (25) datasets, which have varying resolutions. Test accuracies are improved from 74% to 82%, 78% to 93%, and 47% to 88%, respectively.

## Methods

**Physical model of genetically programmable optical random neural network.** Here $A(x,y)$ and $\phi(x,y)$ represent the amplitude and the phase of the incoming laser beam and $S(x,y)$ denotes a sample of a dataset. With the help of a spatial light modulator (SLM), they create the following expression,

$$O(x,y) = A(x,y)\, exp\,[j\phi(x,y)]\, exp\,[jS(x,y)] \quad (1)$$

for the laser beam with the encoded information, $O(x,y)$. The mathematical expression which indicates the operation of the optical random neural network can be shown as

$$R(x,y) = F^{-1}\{F\{O(x,y)\}H(f_x, f_y; \theta)\} \quad (2)$$

$$I(x,y) = |R(x,y)|^2 \quad (3)$$

where $H(f_x, f_y; \theta)$ is the complex, random matrix representing the diffuser (its transfer matrix) as parameterized by the angular position of the disk, $\theta$, $R(x,y)$ is the received optical field at the camera plane, and $I(x,y)$ its intensity recorded by the camera. Here, $x$ and $y$ denote spatial coordinates while $f_x$ and $f_y$ denote spatial frequency coordinates.

The classification operation performed by the readout layer can be expressed as

$$w^* = argmin_w |y_l - Xw|^2 + \alpha |w|^2 \quad (4)$$

where the optimum weights $w^*$ are inferred from a training set with images $X$ and corresponding labels $y_l$. $\alpha$ governs the regularization strength, which is set to 1 throughout all analyses.

**Experimental procedure.** A continuous-wave laser source with a central wavelength of 1064 nm is used for illuminating a spatial light modulator (SLM, HAMAMATSU X10468) with 800 x 600 pixels resolution. Since the samples we used originate from datasets developed for machine learning tasks, they typically have low resolution e.g., 28 x 28 pixels. To be able to maximally use the effective area of our SLM, we performed image upsampling by an integer factor for low-resolution datasets. For datasets with samples having higher resolution than our SLM resolution, we performed image downsampling. The phase encoding process was then carried out as shown by the expressions above.

We placed an adhesive tape at the processing plane as the scattering medium providing random matrix multiplication. The scattering medium was rotated to different angles using a stepper motor (28BYJ-48) to yield parameterizations of the random matrix. Since the classification accuracies depend on the random matrix values, and no a priori relationship is known among different random matrix distributions corresponding to different angles, a heuristic search algorithm, genetic algorithm (GA)(26), was utilized for optimizing for the maximum accuracy.

While programming the optical random neural network, at each GA iteration, all the dataset samples are optically processed with a particular candidate random projection kernel and collected with the camera. All camera parameters, including exposure, saturation, and gain, are kept constant throughout the experiments. Captured 2D RGB images are converted to grayscale images and downsampled by local averaging with a pool size of $(20,16)$ (see Supplementary Discussion 12 for the selection of the pool size). As a final step, downsampled 2D grayscale images are converted to 1D arrays (flattened), and ridge classification is performed. The fitness function in GA is set as the classification accuracy, and depending on the result, the stepper motor is rotated to the angle yielded by GA. Following this pipeline, GA searches and finds the kernel that provides the best accuracy level for designated datasets.

Genetic algorithm (GA) employs biology-inspired operations such as mutation, crossover and selection. In order to utilize GA, we formulated a fitness function where higher accuracies were favored, the input parameter (gene) being angular position specified in terms of the steps of the motor in the interval [0,4095], represented as a bitstring for genetic operations, and the output being the ridge classification accuracy of the randomly-mapped dataset. We specified 12 generations and a population size of 4 for each generation in GA, and explored different GA parameters in Supplementary Discussion 10. In the first generation, all genes were selected randomly, then genes with superior classification accuracies were chosen as mating parent genes (2 in our case) and a new generation was created by crossover (single-point crossover with 80% probability) and mutation (random mutation with 1% probability) operations. The crossover operation randomly selects an index in the bitstrings of the mating parent genes as the crossover point and bits to the right of the crossover point are swapped between the parents, creating a new offspring. The mutation operator, on the other hand, randomly inverts the bits of a gene at certain locations. We obtained the optimal kernel like so. To ensure a nondecreasing nature for the accuracy, we kept the best gene in a given generation and passed it onto the next generation. So for each generation, 3 new genes were chosen by GA. In total, the dataset was passed through 40 locations. Given genetic algorithm parameters where the number of generations is shown with $n$ and the population size with $p$, the total number of random projection kernels that are probed is given by the following formula:

$$n(p - 1) + p \tag{5}$$

To calculate the fitness function corresponding to each gene, we used a laptop with Intel Core i7-7600U CPU and 8 GB RAM that sends commands via serial communication to a microcontroller board (Arduino Uno) which processes step sequences to rotate the stepper motor to the orientation specified by genes. Our experimental procedure, including the hardware flow control and software flow control is schematically illustrated in Supplementary Discussion 4.

## Results

**Numerical studies.** Before constructing the experimental setup, we make use of physically accurate beam propagation method (BPM) simulations to validate our main idea of controlling optical random projection. The details of the simulation are given in Supplementary Discussion 1. Here our design leverages Johnson-Lindenstrauss Lemma(27), which states that random projection can be used for dimensionality reduction purposes(28), which is at the core of learning from high-dimensional data. We benchmark our idea on a binary classification task using the Breast MNIST(23) dataset containing 780 samples, and with an 80/20 train-test split, we achieve the GA evolution plot presented in Fig. 1c. As can be seen from the figure, simulation results show improvements over the baseline ridge classification accuracy of 66.67%, which corresponds to unprocessed samples. After the first iteration in GA, a ridge classification accuracy of 67.31% is reported, which is already above the baseline performance. Based on the classification accuracies in the following iterations, GA proposes an optimal point where the ridge classification accuracy reached its maximum, 77.56%. We present the improvement in classification accuracy by plotting confusion matrices at the beginning (Fig. 1a) and the end (Fig. 1b) of the programming process. The confusion matrices are normalized such that entries on the same row sum up to 100, barring rounding errors. These confusion matrices for the simulated Breast MNIST dataset demonstrate that classification performance can be significantly improved by programming the optical computing platform and searching for an optimal random projection kernel heuristically. After validating the controllability of optical random projection, we move on to the experimental setup and results in the next subsection.

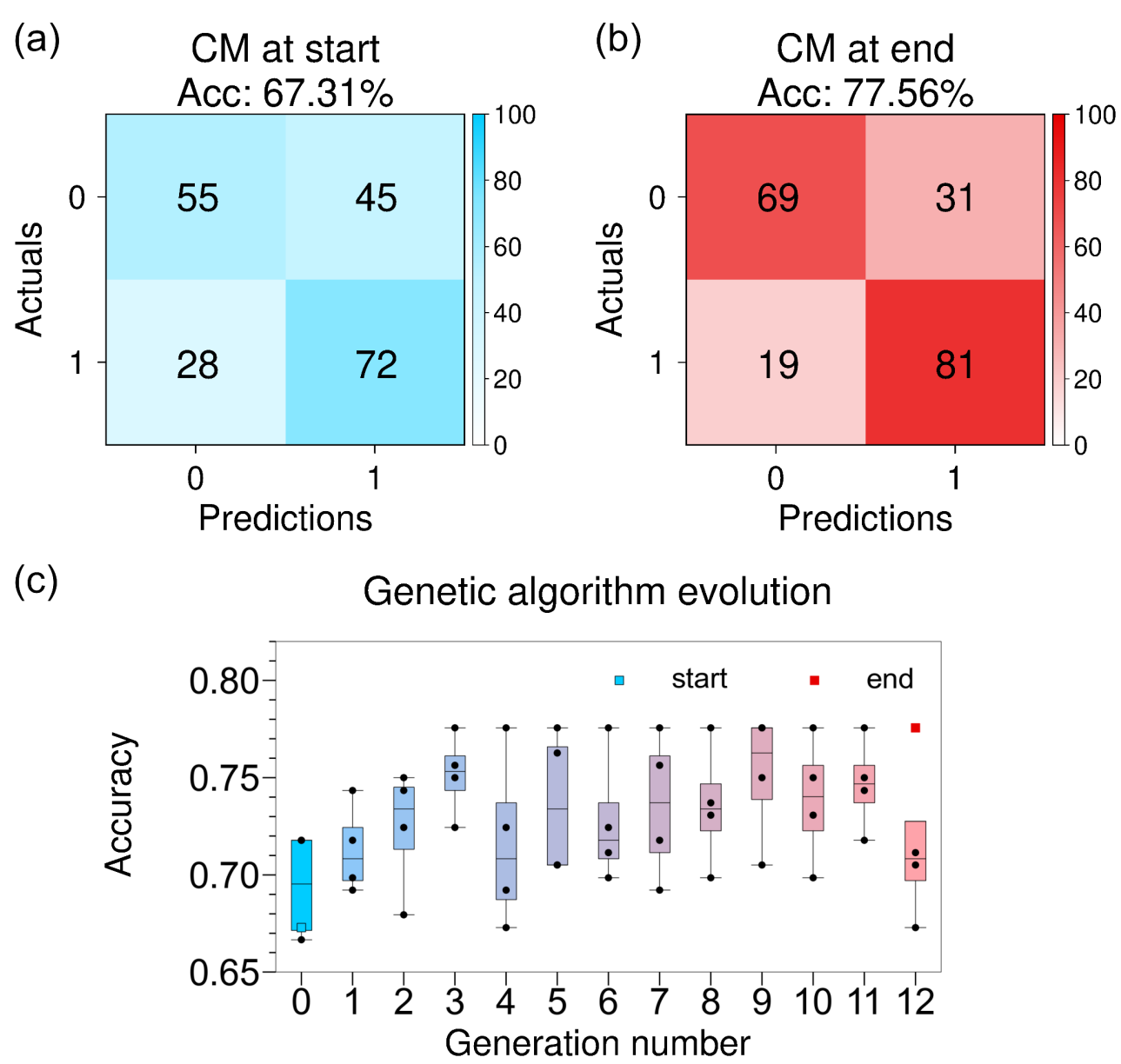

**Figure 1:** Simulated learning results for the Breast MNIST dataset. Confusion matrices (CM's) corresponding to the first (a) and the last (b) genetic algorithm (GA) iteration. (c) Evolution of the accuracy during GA.

**Experimental studies.** Experimental realization of our genetically programmable optical random neural network is illustrated in Fig. 2. It consists of a continuous-wave (CW) laser source, a spatial light modulator (SLM), a pair of lenses to perform Fourier transforms, a scattering medium placed on a disc located on the Fourier plane to provide optical random projection and a camera to collect optically processed information. The optical information encoding process is carried out by superimposing each sample of the dataset, an 8-bit image, as a phase pattern on the SLM. After the encoding step, the information-carrying optical field is processed linearly all the way up until the recording plane. Optical random projection/mapping is realized with a diffuser.

When coherent light passes through the diffuser, different optical free paths are formed at different points on the scattering media, and at the camera plane, we obtain a speckle pattern. A small change in the orientation of the diffuser changes the result of the element-wise matrix multiplication, i.e., the convolution kernel is dependent on the orientation of the diffusing medium. Therefore, we propose searching for and finding an optimum kernel (diffuser surface) with better feature extraction capabilities. Since the information-carrying laser beam occupies a small region on the diffuser, introducing a new region to it by rotating the diffuser changes the random projection kernel. For this purpose, we constructed a disk covered with adhesive tape, which resulted in a search space where there is only one degree of freedom (angular position), allowing us to search for better kernels easily.

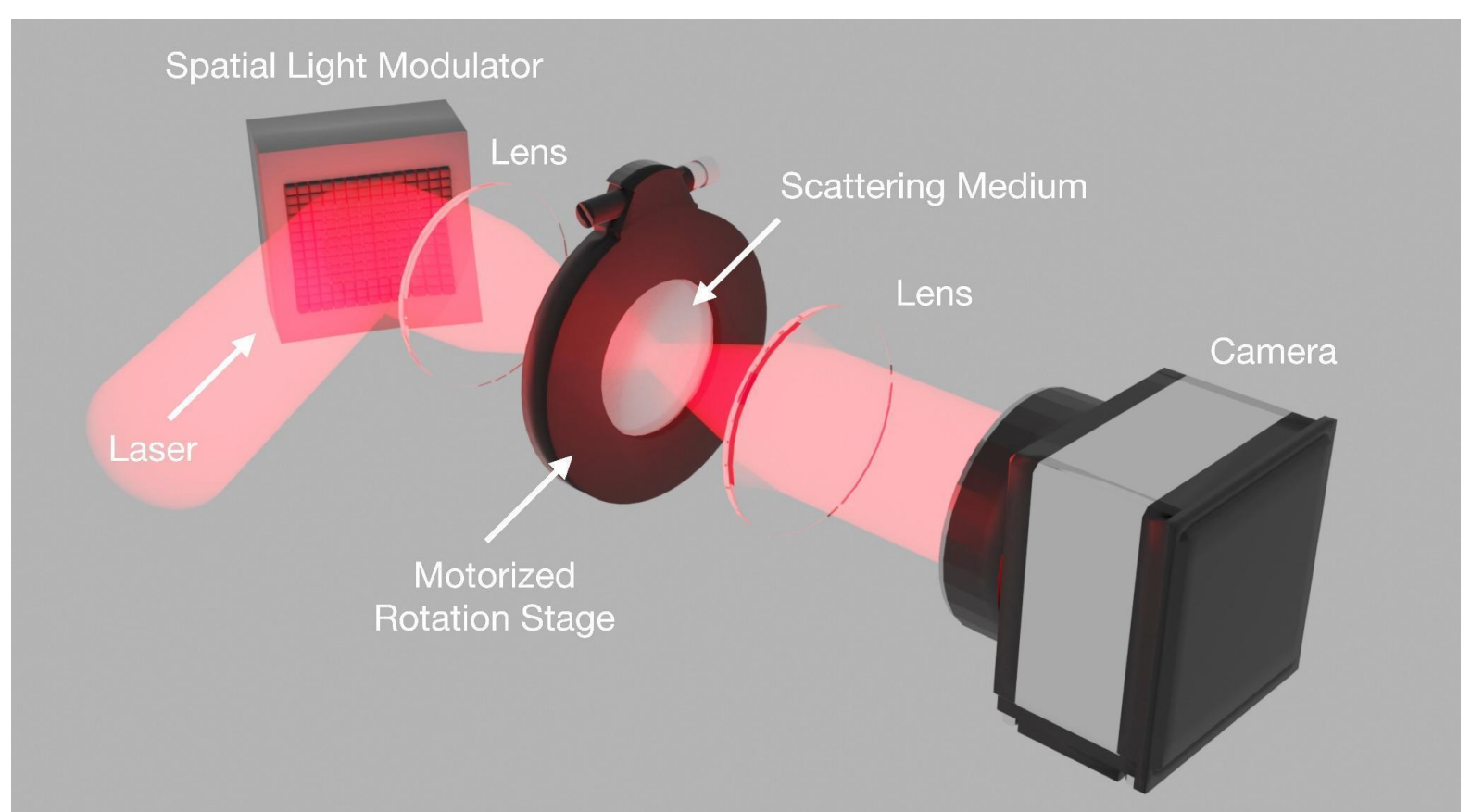

**Figure 2:** Schematic of random projection-based programmable optical computing setup. The experimental setup consists of a spatial light modulator for encoding information onto laser light, a scattering medium on off-axis rotation stage, and an imaging system to relay light and a camera to decode information.

A stepper motor is employed to control the orientation of the diffuser with the half-stepping method, where one full rotation of the disk is divided into 4096 steps. Without benefitting optimization algorithms and trying each possible kernel individually, evaluating the resulting accuracy would take a long time experimentally. Thus, we utilize a heuristic search algorithm, genetic algorithm (GA)(26), for optimizing for the maximum accuracy, which is a function of the angular position, $\theta$. Such a method requires several trials to meet the designated task and decreases optimization time in experiments. The deployment of a heuristic search algorithm like GA can also be attributed to the uneven structure of the complex media, where there is no a priori relationship between the complex random matrices obtained from distinct points significantly far from each other. However, we want to emphasize that since the rotation steps are small, similarities in output images corresponding to matrices originating from a neighborhood of consecutive angular positions are observed in our experiments (see Supplementary Discussion 5 for quantitative metrics). GA parameters used in this study and the backend processing steps are explained in Methods.

In experiments, we initially employ GA on the Breast MNIST(23) dataset and obtain the GA evolution plot shown in Fig. 3c. With the same 80/20 train-test split as in simulations, the ridge classification accuracy is improved to 82.05% starting from 73.72%. The confusion matrices corresponding to the first and the best GA iteration are given in Fig. 3a and b, respectively. To further ensure the claimed improvement is not due to noise, we perform 5-fold cross-validation on randomly projected samples corresponding to the best GA iteration. As a result, a mean classification accuracy of 73.08% with a standard deviation of 7.95% was obtained. For comparison, the first GA iteration produces 5-fold cross-validation metrics with a mean classification accuracy of 71.28% with a standard deviation of 2.88%. We additionally performed paired-sample Student's t-test and the returned value of $h = 1$ indicates that the t-test does reject the null hypothesis at the 1% significance level.

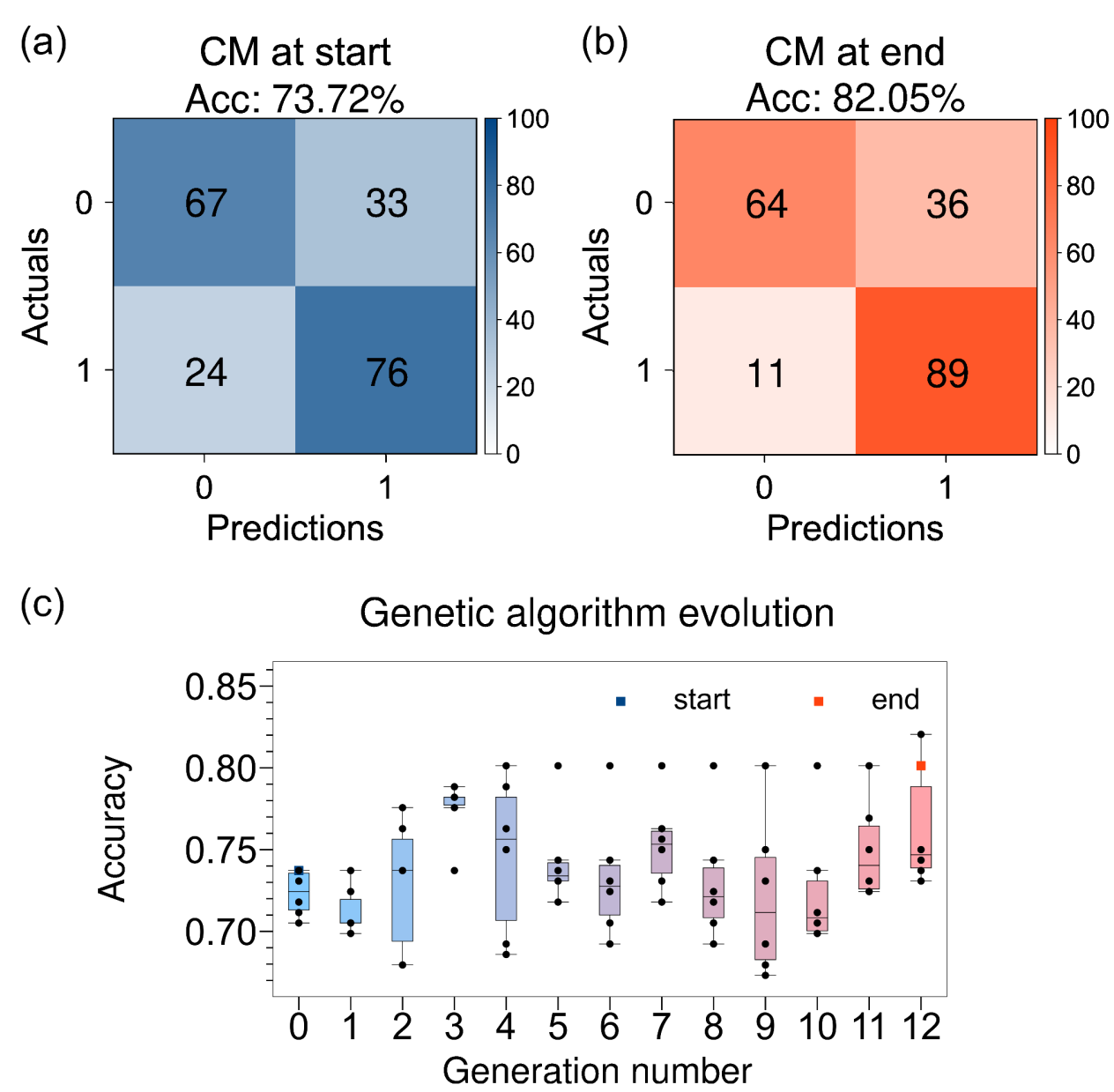

**Figure 3:** Experimental learning results for the Breast MNIST dataset. Confusion matrices (CM's) corresponding to the first (a) and the last (b) genetic algorithm (GA) iteration. (c) Evolution of the accuracy during GA.

After experimentally demonstrating and validating the tunability of optical random projection on a small-scale dataset with samples having 28 x 28 pixels resolution, we note that we can accommodate high-resolution samples in our experimental setup and fully leverage the parallelism provided by our optical computing platform, where we are only limited by the SLM resolution, which is 600 x 800 pixels. For this reason, we decided to scale up and tackle a more complex classification task using the COVID-19 X-Ray(24) dataset with higher resolution (400 x 400 pixels) samples. However, the SLM used in this study operates at a 60 Hz repetition rate since liquid crystal technology is limited in terms of speed. To decrease the time spent on the programming of the random projection kernel and make our optical computing method useful for larger datasets, we propose to form a smaller subset of the dataset to be considered a proxy to the full set for the programming step. For this purpose, out of the full dataset (2481 samples), a subset of size 300 containing randomly selected 150 positive and 150 negative samples is created. As previously, an 80/20 train-test split is used. This way, our goal is to reduce the programming time while maintaining the improvements in classification accuracy. As demonstrated in Fig. 4b, we observe a significant increase in the ridge classification accuracy from 78.33% to 93.33% while programming the optical computing platform with GA.

To evaluate the performance of the subset method, we optically process all the samples of the COVID-19 X-Ray dataset with the optimized random projection kernel proposed by GA, employing an 80/20 train-test split on the whole dataset. An accuracy of 90.14% is achieved over the baseline accuracy level of the dataset, which is 74.85%. These results demonstrate that the genetically programmed optical computing platform provides 15% higher accuracy. In contrast, when we use GA for the full dataset instead of 300 samples, we observed a maximum accuracy level of 91.15%. It shows that by utilizing a randomly selected subset and decreasing the programming time significantly, we sacrifice a 1% performance difference in classification, and even a small subset gives rise to improvements in classification accuracy. We can observe the evolution of accuracy via the confusion matrices presented in Fig. 4a and c. The confusion matrix corresponding to the entire dataset is also given in Fig. 4d. With the COVID-19 X-Ray dataset results, we validate the high-resolution sample processing capability of optical random projection on top of programmability demonstrated earlier. We would like to note that such high resolutions are atypical in conventional machine learning frameworks.

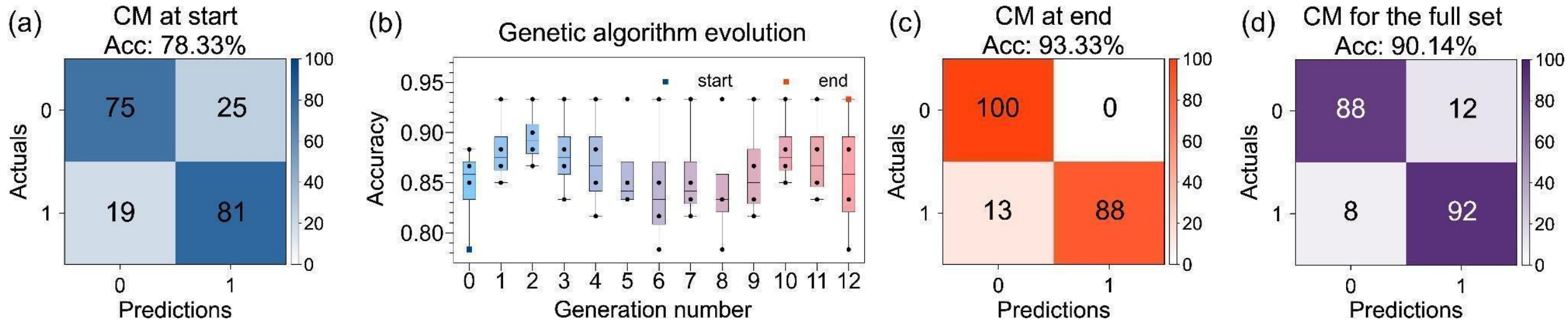

**Figure 4:** Experimental learning results for the COVID-19 X-Ray dataset when a subset of the full dataset is used for genetic algorithm (GA). (a) Confusion matrix (CM) corresponding to the first GA iteration. (b) Evolution of the accuracy during GA. (c) CM obtained at the end of GA. (d) CM corresponding to the full dataset when all the samples are passed through the optimal angular position yielded by GA.

After experimentally validating the programmability and high-resolution sample processing capabilities of our scheme, we explore the retinal disease classification dataset (25) containing 1493 training and 439 test samples with 1424 x 2144 pixels resolution. The images are classified into two classes based on whether any type of anomaly is present or not. The samples are first converted from RGB to grayscale and image downsampling with a constant factor is performed to effectively use our SLM having less pixels than samples. In other words, downsampled images are transformed to have 600 x 800 pixels resolution, which is our SLM resolution. However, we note that with next-generation SLMs having 4K resolution, such downsampling can be avoided to process high resolution images. The remaining hardware & software settings are the same as before. Fig. 5 shows the evolution of the accuracy during GA and the confusion matrices for this dataset. After programming the kernel, we observe an increase in classification accuracy from 46.70% to 87.70%, which is better than the performance of a digital ridge classifier, which gives 85.65% accuracy with a similar number of parameters.

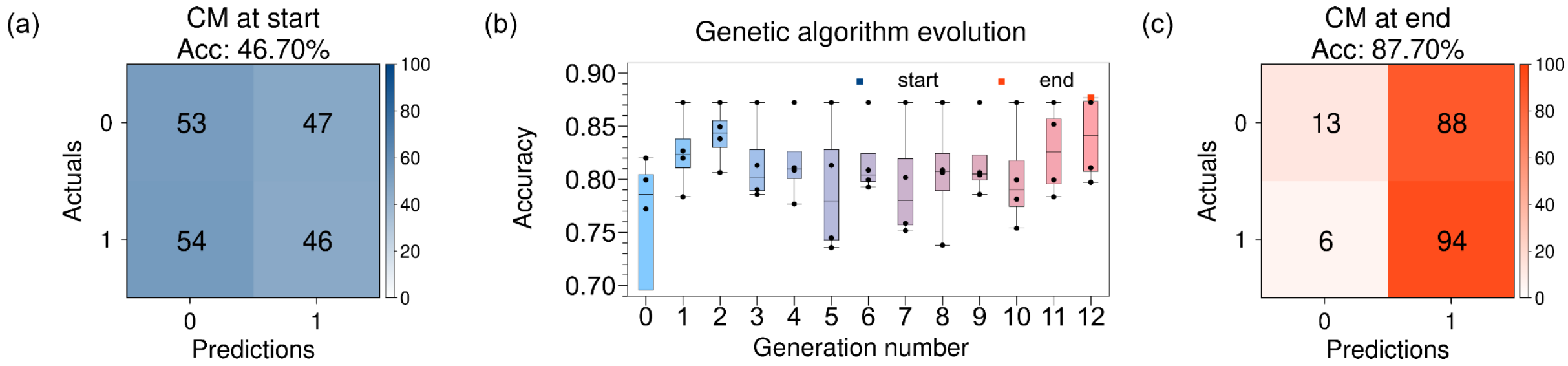

**Figure 5:** Results for the high resolution retinal disease classification dataset. (a) Confusion matrix (CM) corresponding to the first GA iteration. (b) Evolution of the accuracy during GA. (c) Confusion matrix obtained at the end of GA.

We would like to emphasize that the imbalance in confusion matrices is due to the bias in the dataset, where both the training and the test set are dominated by positive samples, and would like to focus on the overall classification accuracy. Concretely, the dataset consists of 1674 positive samples and 258 negative samples. It must also be noted that we obtained out-of-memory errors when we tried to train a vanilla CNN having the same inputs as the optical platform (samples with 600 x 800 pixels resolution) on an NVIDIA RTX 4070 graphics card with 12 GB of RAM. Therefore, our proposal being resolution-agnostic is a major advantage over conventional techniques, where high-resolution samples typically need to be downsampled. Additional results on high-resolution geospatial datasets can be found in Supplementary Discussion 9.

## Discussion

To understand the effect of the programming random projection kernel with a genetic algorithm, we perform linear discriminant analysis (LDA), a dimensionality reduction technique, on one of the widely-used datasets for benchmarking different machine learning architectures, which is the Fashion MNIST (29) dataset. We feed randomly projected samples to LDA, which helps us to illustrate the inner workings of our optical computing platform. In Fig. 6, 960 features obtained from the randomly mapped subset (3000 samples) of the Fashion MNIST dataset are represented by three components. The distribution of color-coded classes shows that the clustering performance of optical random projection improves as GA progresses. The clustering of similar samples eases the task of the readout layer, thus increasing the performance of the optical computing method. We would like to emphasize that Fashion MNIST samples have low resolution (28 x 28 pixels) and do not reflect the high-resolution sample processing capability of our programmable optical random projection approach, which is why we have left experimental learning results to Supplementary Discussion 8.

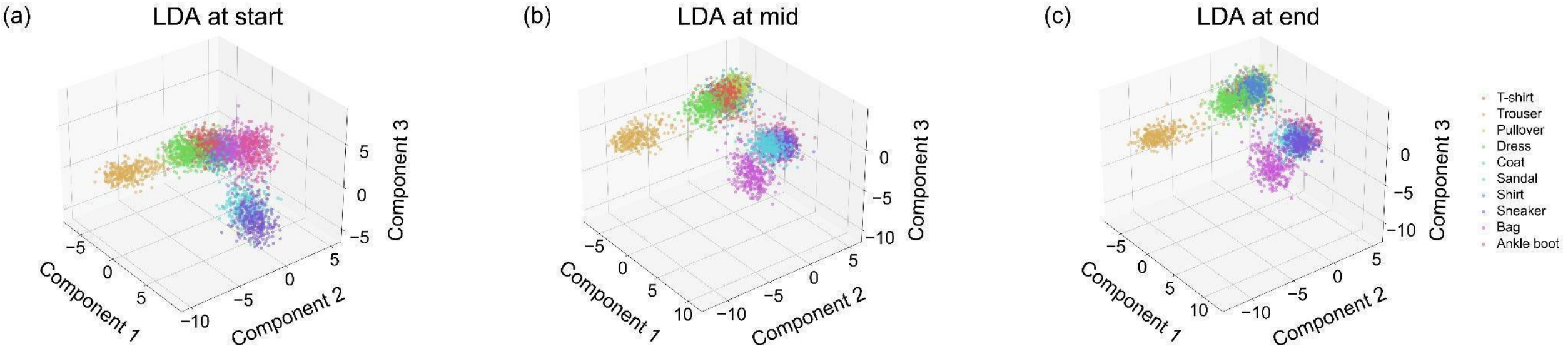

**Figure 6:** Linear discriminant analysis (LDA) performed on a subset of the Fashion MNIST dataset (a) before genetic algorithm (GA), (b) for an arbitrary iteration of GA, and (c) after GA has found the optimal random projection kernel.

Without sacrificing the simplicity of the demonstrated optical computing platform, a nonlinear activation function such as the inverse rectified linear unit (ReLU) function can be implemented by partially saturating the measured beam profiles via either increasing the laser power or tuning the camera parameters. In the literature, the effect of such an additional nonlinearity has been reported(30). It can increase the accuracy of the machine learning tasks when incorporated with the absolute square measurement introduced by the intensity detection of the camera at the readout layer.

In our study, the preferred camera can record 8-bit grayscale images. However, we emulate a camera capable of representing images using fewer bits to test our approach in data-scarce settings by mapping the range 0-255 to a smaller range. We observe that with such a device, faster data rates would be enabled without sacrificing the classification accuracy (see Fig. S13 and Table S1 in Supplementary Discussion 12). We conclude that with suitable hardware tailored for modern machine vision tasks requiring real-time performance, our programmable reservoir can continue to perform machine learning tasks with high accuracy.

The presented programmable optical computing platform can be further improved with alternative programming algorithms. For example, one can use surrogate optimization tools(31) instead of GA, choose a different/learnable weighting parameter $\alpha$ in Eq. 4 during ridge classification, or change the pool size used in local averaging, which would affect the number of features pertaining to each captured image. Although we have yet to opt to make these changes for the sake of simplicity and not losing the essence of our endeavor, they can potentially increase the ridge classification accuracy of the readout layer. In addition, our approach can be augmented by using the recurrence relations of the reservoir.

In all experiments, phase objects were processed and classified. However, amplitude objects and, more generally, incoherent data can be processed using our approach as well. In this case, the speckle response will be different than the coherent case because constructive/destructive interference will not occur under broadband illumination. Since the irradiances will only be superposed without the contribution from cross-terms, the complexity of the system will be lower, and the feature extraction capabilities of the incoherent setup will be affected.

Broader applicability of our method can be realized by imposing gratings in x- and y-axes on input samples, where this time, instead of angular position, the grating x and y values are to be optimized. Therefore, programmable optical random projection in our optical computing platform without any moving parts can be constructed in principle.

Energy consumption in our setup is only due to the power requirements of components such as the laser source, SLM, and camera. All processing, including the random matrix multiplication, is done passively and our computing platform relies on linear optics. It indicates that the consumed energy is independent of the sample resolution. Energy consumption in our method and conventional electronics is further compared in Supplementary Discussion 14. Our work opens up an avenue for energy-efficient, resolution-agnostic and programmable optical processing schemes for success in modern machine learning practice.

**Funding**

This work is supported by the Scientific and Technological Research Council of Turkey (TÜBİTAK) under grant number 123E308.

**Acknowledgements**

We thank Aydoğan Özcan for fruitful discussions and Çağrı Şenel for providing the SLM and laser source used in experiments.

**Data availability statement**

Data and code related to the results in this work may be obtained from the authors upon reasonable request.

**Conflicts of interest**

The authors declare no conflicts of interest.

**Supplementary information** The online version contains supplementary material available at …

# Supplementary information for

# Genetically programmable optical random neural networks

Bora Çarpınlıoğlu, Uğur Teğin*

Department of Electrical and Electronics Engineering, Koç University, Istanbul, 34450, Türkiye
*utegin@ku.edu.tr

### Supplementary Discussion 1: Breast MNIST Simulation Details

Prior to conducting any experiments, we first validated our approach with beam propagation method (BPM) simulations. Since our experimental design is based on linear operations, we solve numerically the linear wave propagation equation. Working in the linear domain allows us the ability to perform one-step simulations i.e., the propagation does not need to be divided into small steps of dz.

In line with our experimental prototype, we defined a spatial grid covering the spatial light modulator (SLM) active area where modulation is performed. A Gaussian beam is created with samples from the Breast MNIST dataset encoded as phase objects. Since the samples in the dataset have small (28 x 28 pixels) resolution, they are upsampled to fit the active area of the SLM. Following linear beam propagation and focusing with a lens, the complex field is multiplied with one of the 4096 pre-computed random matrices, determined by genetic algorithm (GA). The random matrices are defined as follows: an initial, fully random matrix is generated corresponding to position 0. Then, the random matrix at position 1 is created with half of the elements originating from the random matrix at position 0, and the remaining half is created randomly. This procedure goes on until the last random matrix. As a result, consecutive random matrices become correlated, emulating the stepper motor rotation in the experimental setup. After multiplication with a given random matrix, normalized intensity profiles are obtained with another relaying lens and linear beam propagation, after which ridge classification is performed to obtain classification accuracy. In simulation, GA parameters of $n = 12$ generations and $p = 4$ population size are used. An example BPM script can be found on GitHub (1).

### Supplementary Discussion 2: 4f Imaging System and Input Encoding

The beam expander with the Keplerian design applied to the laser beam consists of two lenses placed in parallel at a distance equal to the sum of their focal lengths and its magnification can be calculated by the ratio of the focal lengths of the constituent lenses. The lenses used in this 4f imaging system are two bi-convex lenses with focal lengths of 30 mm and 268 mm. In the 4f imaging system, input beam inversion does not pose a problem for us because the output beam has the same intensity pattern as the input due to the azimuthal symmetry of the input Gaussian beam. As a result, SLM receives a magnified laser beam whose input intensity pattern did not change. Similarly, after reflection from the SLM, we use a 4f imaging system, again using two bi-convex lenses with focal lengths of 268 mm and 50 mm.

In the encoding step, since the size of the samples in the datasets was often smaller than the size of the SLM, we performed image upsampling using bilinear interpolation for each sample such that the illuminated region on the SLM was filled with the sample. In areas where the SLM could not be filled by the rescaled image, we simply performed zero-padding.

**Supplementary Discussion 3: The Experimental Setup and the Scattering Medium Used in Experiments**

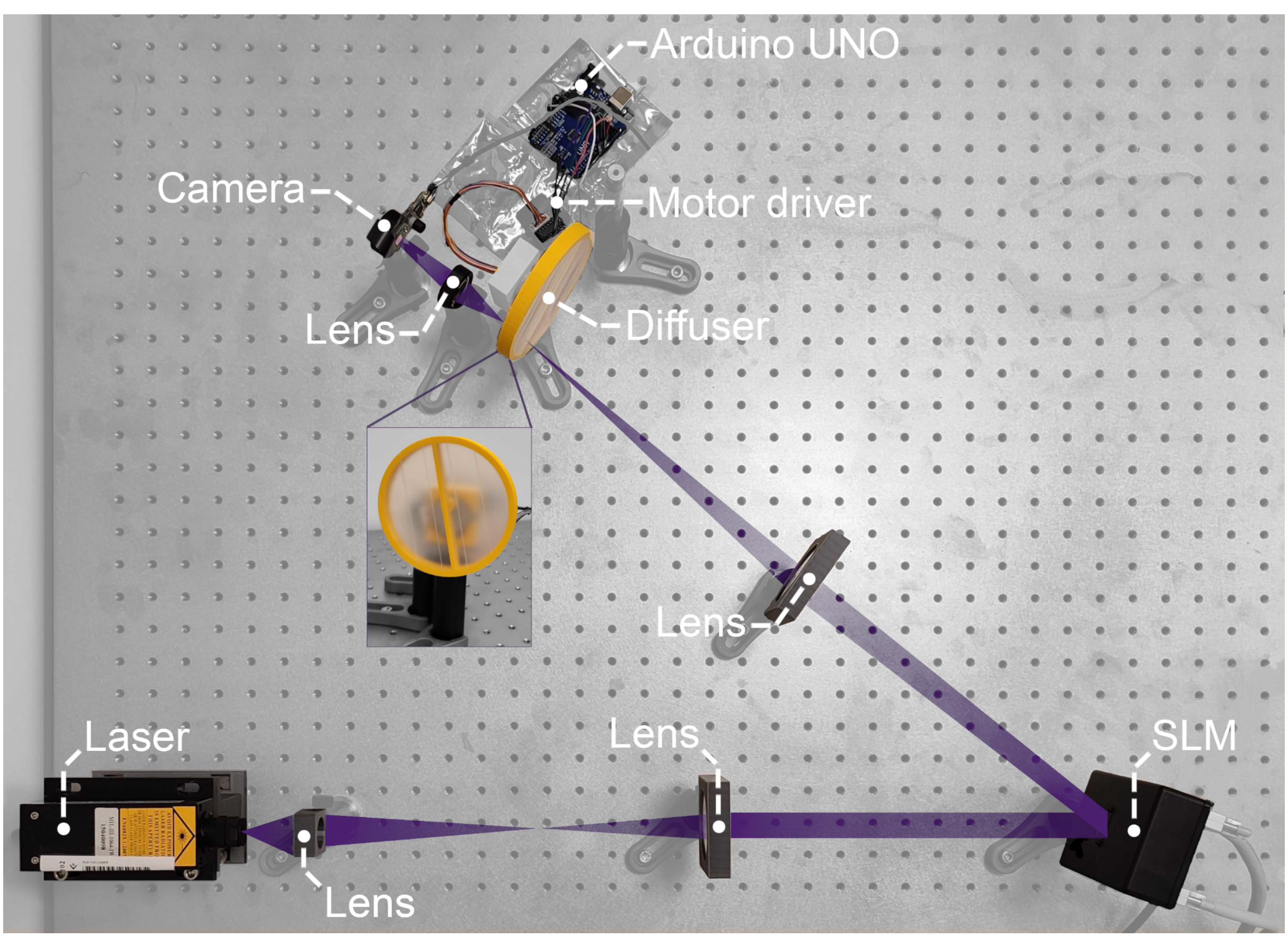

**Figure S1:** The experimental setup used for genetically programming random projection kernels. A disk covered with adhesive tape is used as the scattering medium, as seen on the inset. SLM: Spatial light modulator.

## Supplementary Discussion 4: Hardware Flow Control and Software Flow Control

In Fig. S2 we illustrate our computing platform as a diagram including the interaction between the optical and electrical layers and the flow of our genetic algorithm-based scattering medium optimization.

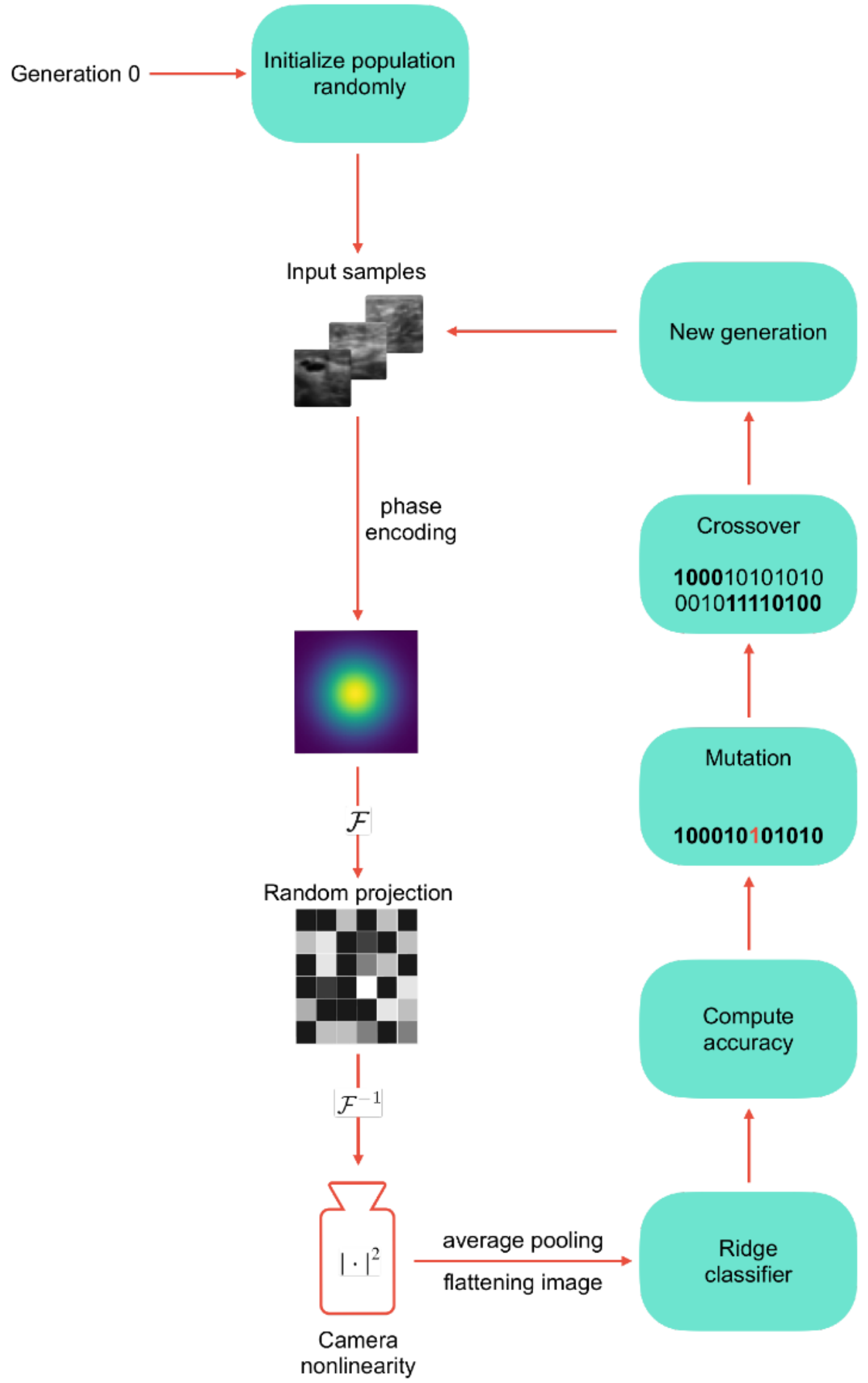

**Figure S2:** The hardware and software flow control of our proposal. The optical steps are illustrated with images and the software part is illustrated in light blue. Genetic algorithm operations such as mutation and crossover are shown schematically.

## Supplementary Discussion 5: Similarity Between Adjoint Random Projection Kernels

To see the similarity between different random projection kernels, we performed two additional experiments. First, we fixed a reference point on our diffuser (disk with the adhesive tape) and rotated it within a range of 9 steps in either direction, which corresponds to an angular span of $1.58°$, and recorded the resulting image at each angular position. We hypothesized that, by calculating the similarity between any image and the reference image, we could get a measure of the relation between corresponding random projection kernels. Because of its popularity in image processing and deep learning, we used the structural similarity index measure (SSIM). Fig. S3 shows the average SSIM for five trials. As can be seen in the figure, the general trend manifests itself as a negative correlation between angular displacement (in absolute terms) and similarity.

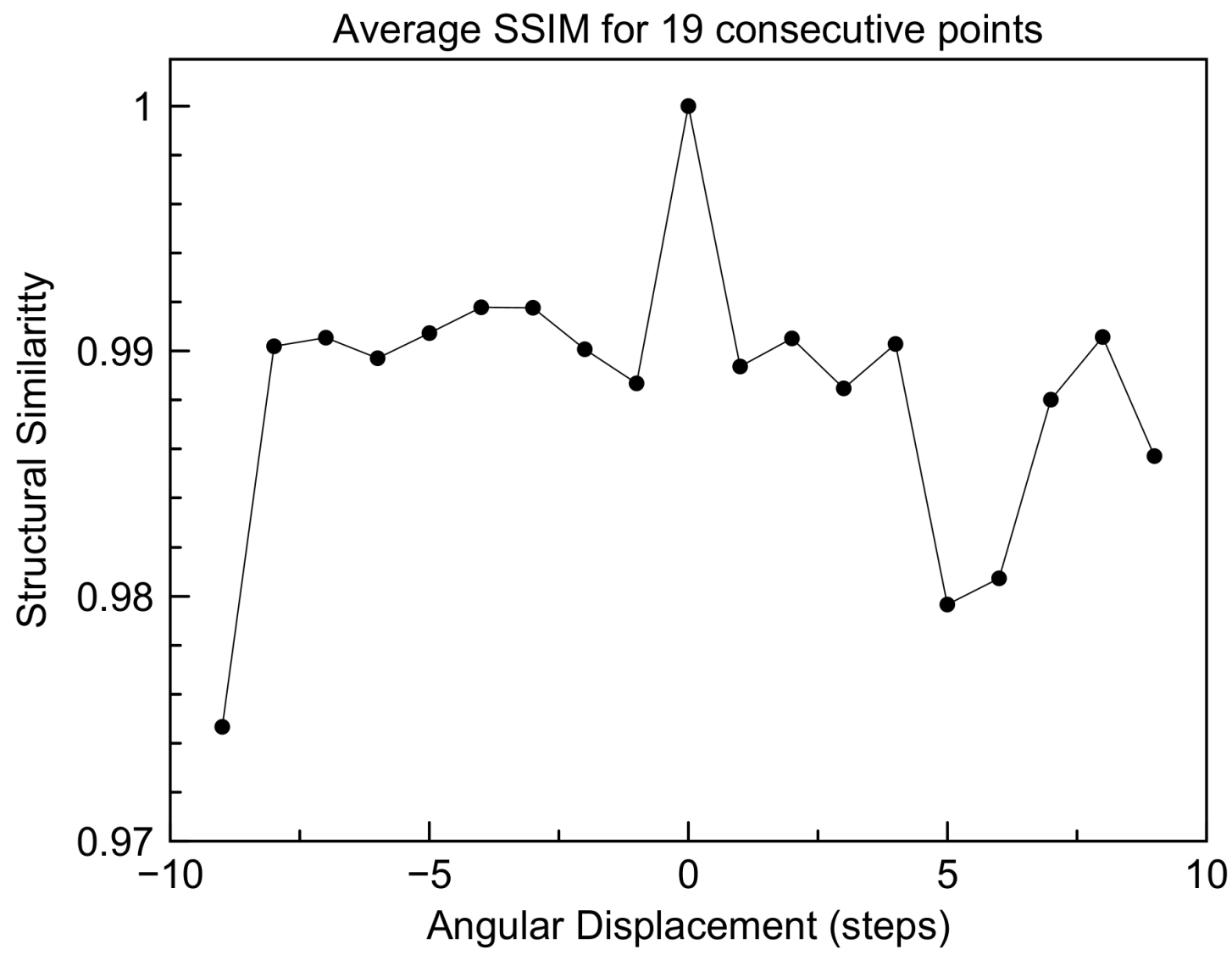

**Figure S3:** Average structural similarity index measure (SSIM) for consecutive points on the diffuser.

Second, we rotated the diffuser for $0°$, $90°$, $180°$ and $270°$ while a sample from the Fashion-MNIST dataset is displayed on the SLM. In Fig. S4 we show captured images and normalized average difference images together with SSIM values when the no rotation case is set as a reference.

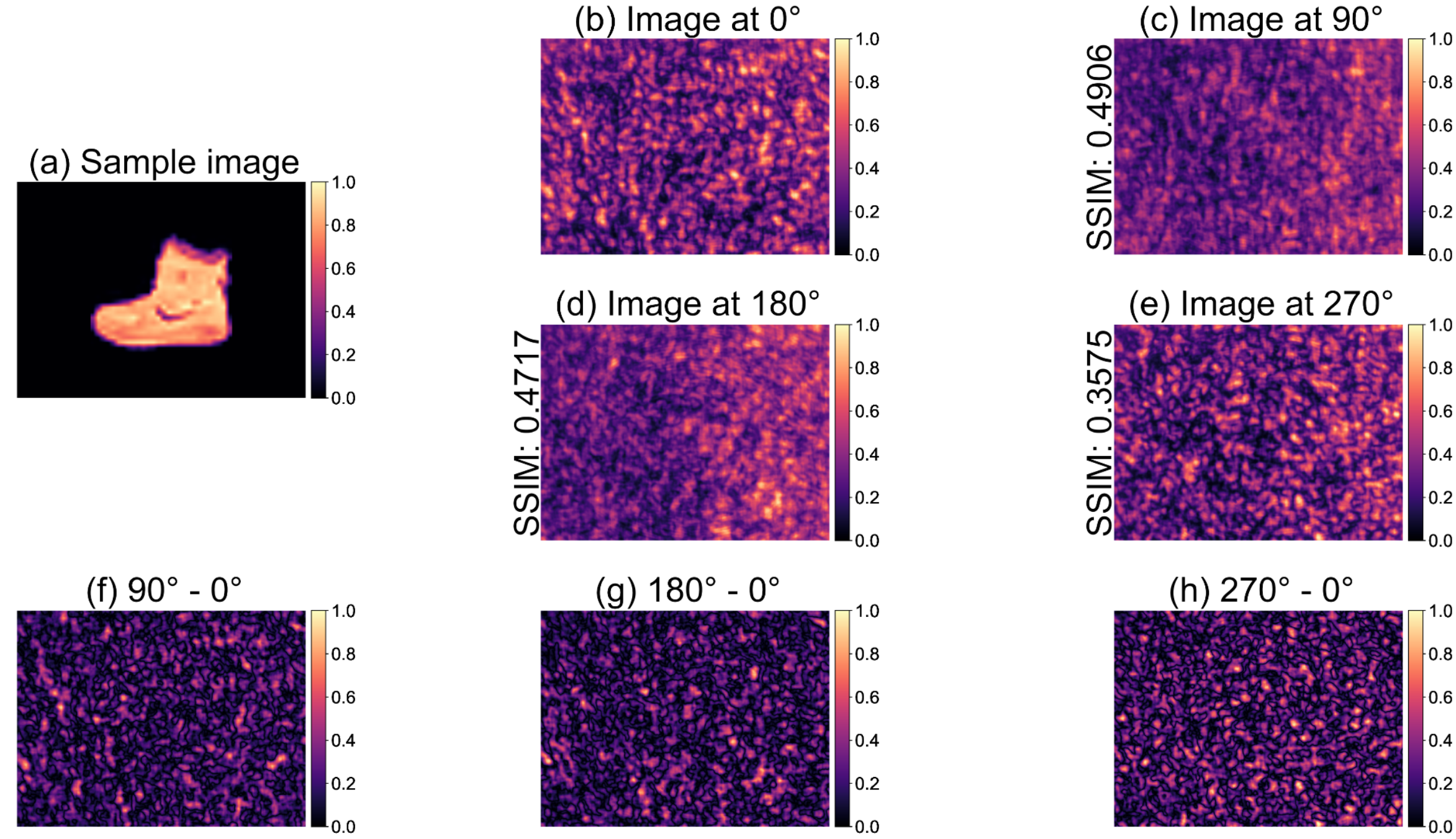

**Figure S4:** (a) Sample image from the Fashion MNIST dataset. (b-e) Captured images when the diffuser is rotated for $0^\circ$, $90^\circ$, $180^\circ$ and $270^\circ$. Structural similarity index measure (SSIM) values are given to the left of the images when applicable. (f-h) Difference images when the no rotation case is considered as a reference.

**Supplementary Discussion 6: Effects of Processing Input Images with Consecutive Random Projection Kernels**

Fig. S5 shows captures corresponding to processing the same input image with 7 consecutive random projection kernels.

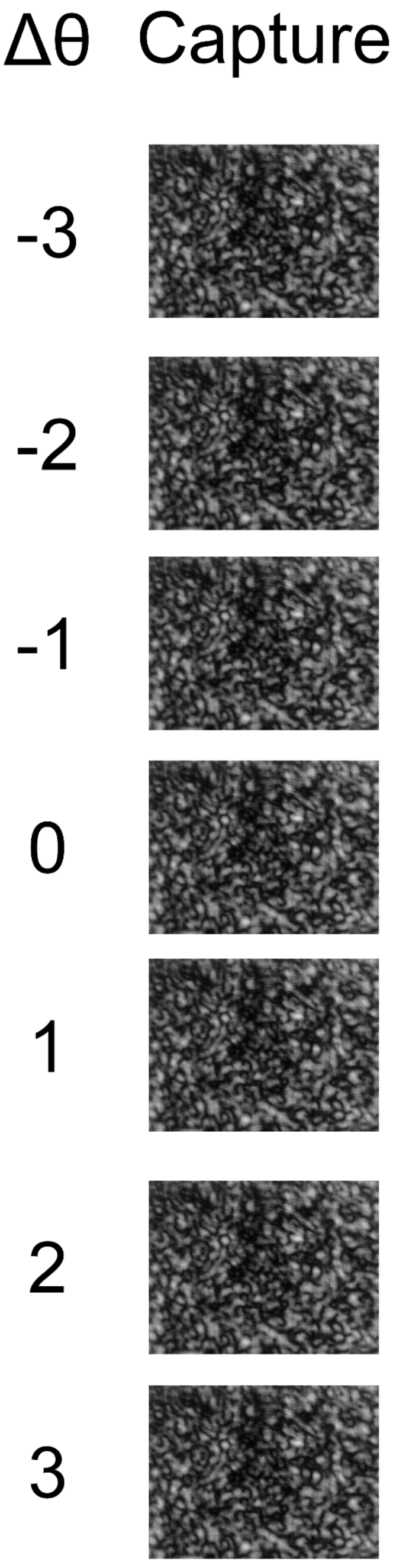

**Figure S5:** Captures corresponding to 7 consecutive random projection kernels. $\Delta\theta$ shows the difference in stepper motor units when the zero case is considered as a reference. Positive and negative $\Delta\theta$ values correspond to clockwise and counterclockwise rotation, respectively.

## Supplementary Discussion 7: Further Results on the Subset Method

As discussed in the main text, accuracy results related to the COVID-19 X-Ray dataset were obtained using a subset method. We also performed GA on the full dataset to be able to assess the effectiveness of the subset method i.e., when all 2481 samples are randomly mapped. For the whole dataset, we obtained an accuracy of 91.15% compared to 90.14% obtained from 300 samples, which shows the approximation capability of the subset method. Fig. S6 shows the confusion matrices and the evolution of the accuracy in this setting.

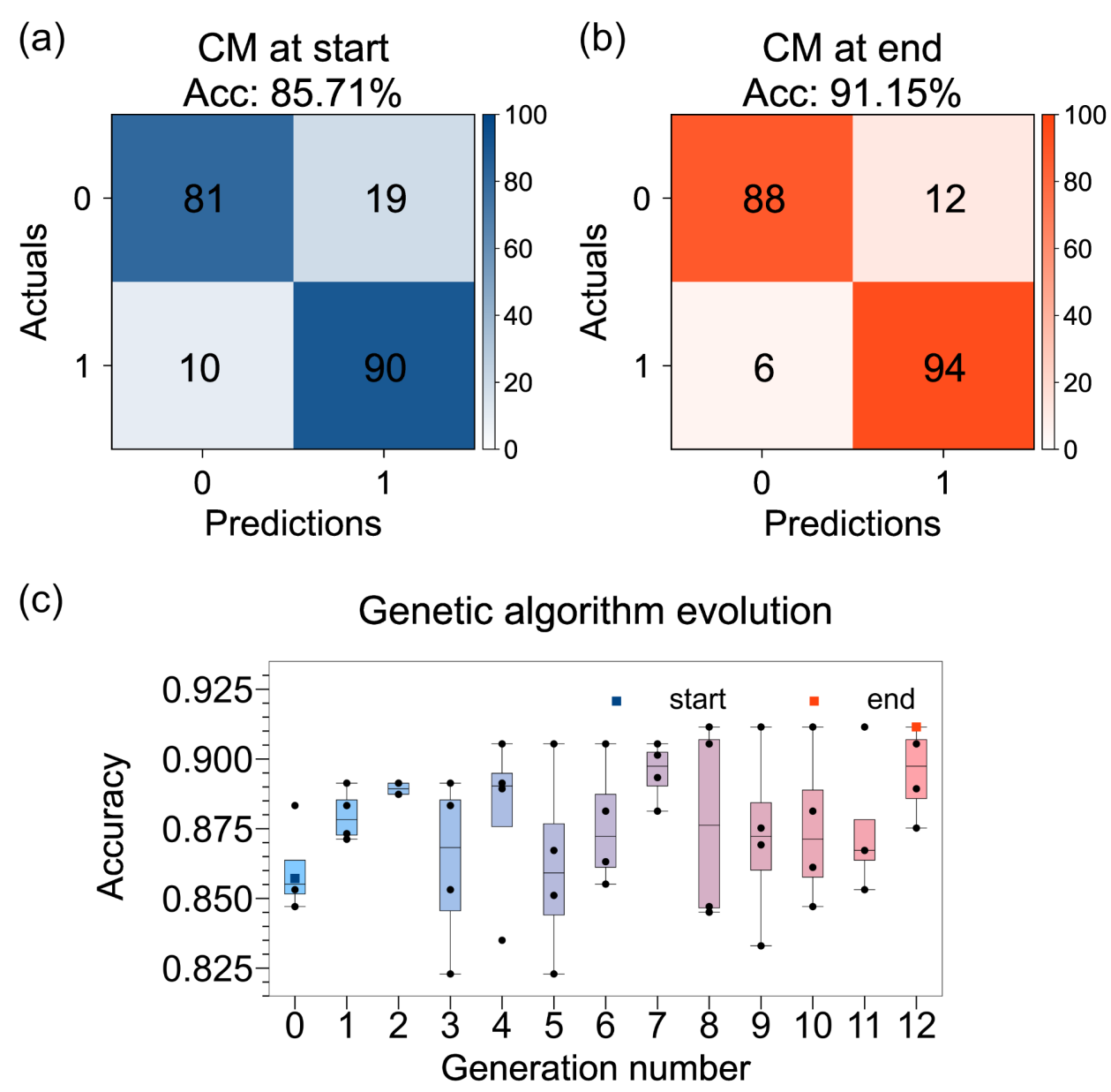

**Figure S6:** Results for the COVID-19 X-Ray dataset when the full set is used for genetic algorithm (GA). (a) Confusion matrix (CM) corresponding to the first GA iteration. (b) Confusion matrix obtained at the end of GA. (c) Evolution of the accuracy during GA.

## Supplementary Discussion 8: Performance on the Fashion MNIST Dataset

We have also tested the popular Fashion MNIST dataset with 70,000 samples(2) to benchmark the performance of our programmable optical random projection scheme and to evaluate our method's performance for multilabel classification. Similar to the COVID-19 X-Ray dataset, we utilize generating a subset to decrease the programming time of our optical neural network. Thus, a subset of randomly selected 3000 samples with an 80/20 train-test split is created from 70,000 samples of the Fashion MNIST dataset. To set a baseline for our method, we apply the ridge classification over the selected subset and obtain 73.83% accuracy. At the beginning of the programming, after the first GA iteration, an accuracy of 71.33% is obtained. At the end of the programming, this classification accuracy is increased to 81.00% (Fig. S7b), resulting in an approximately 10% improvement. The entire Fashion MNIST dataset with 60,000 training and 10,000 test samples is optically processed for the programmed condition, and the classification accuracy of 83.06% is achieved. The row-wise normalized confusion matrices captured during the evolution of GA (Fig. S7a, c) and inference (Fig. S7d) further supported our claim that programming random neural networks effectively decreases classification errors.

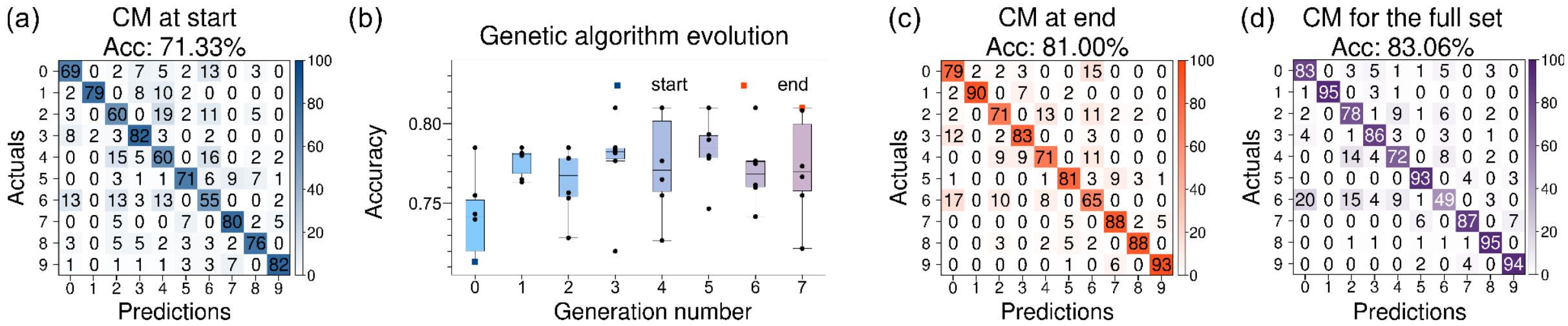

**Figure S7:** Experimental learning results for the Fashion MNIST dataset when a subset of the full dataset is used for genetic algorithm (GA). (a) Confusion matrix (CM) corresponding to the first GA iteration. (b) Evolution of the accuracy during GA. (c) CM obtained at the end of GA. (d) CM corresponds to the full dataset when all the samples are passed through the optimal angular position yielded by GA.

## Supplementary Discussion 9: Performance on High Resolution Multilabel Classification Problems

To better support our approach in terms of scalability in resolution, we perform additional multilabel classification experiments with the RS_C11 dataset (6) with 1232 samples each having 512 x 512 pixels resolution and the Aerial Image Dataset (AID) (7) having 10,000 samples each having 600 x 600 pixels resolution, both of which fit in our SLM. The task is to classify satellite images based on the scene categories they represent. There are 11 and 30 categories for RS_C11 and AID, respectively. An 80/20 train/test split is used for both datasets.

Fig. S8 shows the evolution of classification accuracy during GA iterations for both datasets, where confusion matrices are not visualized because of the high number of classes. Without any optical processing, the RS_C11 dataset yields 12.96% accuracy when ridge classification is solely used. The first GA iteration then yields 10.93%, which is further optimized to be 21.05%. The results for the RS_C11 dataset show the superiority of our approach compared to ridge classification.

For AID, we employed the same subsetting scheme that was discussed for the COVID-19 X-Ray and the Fashion MNIST datasets because of the high number of samples. In other words, we collected a subset consisting of 3000 samples (100 samples per scene category) and let GA run to optimize this subset. As previously, before any optimization was performed, the ridge classification accuracy of the subset was found out to be 25.83%. At the first GA iteration, the optically processed subset produced an accuracy of 22.33% which is worse than the unprocessed ridge classification accuracy. However, as can be seen from Fig. S8b, we ended up with 34.83% accuracy. After completing GA iterations, we used the optimized random projection kernel to obtain 25.95% accuracy compared to the unprocessed accuracy of 12.9% for the whole dataset. Further comparisons for both datasets are given in Table S2.

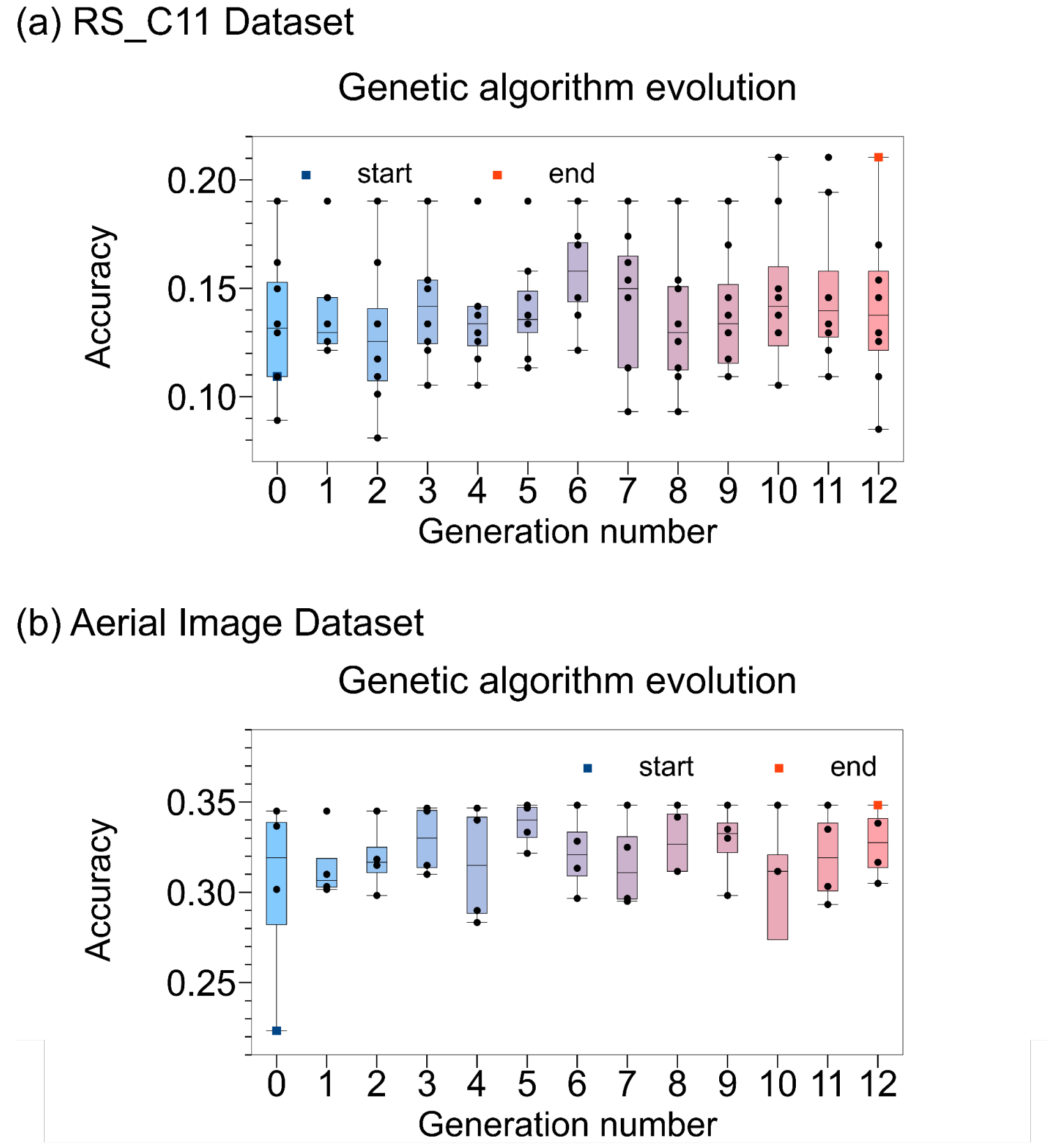

**Figure S8:** Evolution of classification accuracy for the high resolution multilabel geospatial (a) RS_C11 and (b) Aerial Image datasets.

## Supplementary Discussion 10: Effect of GA Parameters on Classification Accuracy

In this section, we investigate the effects of the number of generations and population size on classification performance for some of the datasets evaluated in the main text. Towards this end, for each dataset, we perform multiple experiments where the number of generations is fixed at $n = 12$ and the population size is swept as $p = 4, 6,$ and $8$. Fig. S9 plots the accuracy versus population size for this setting.

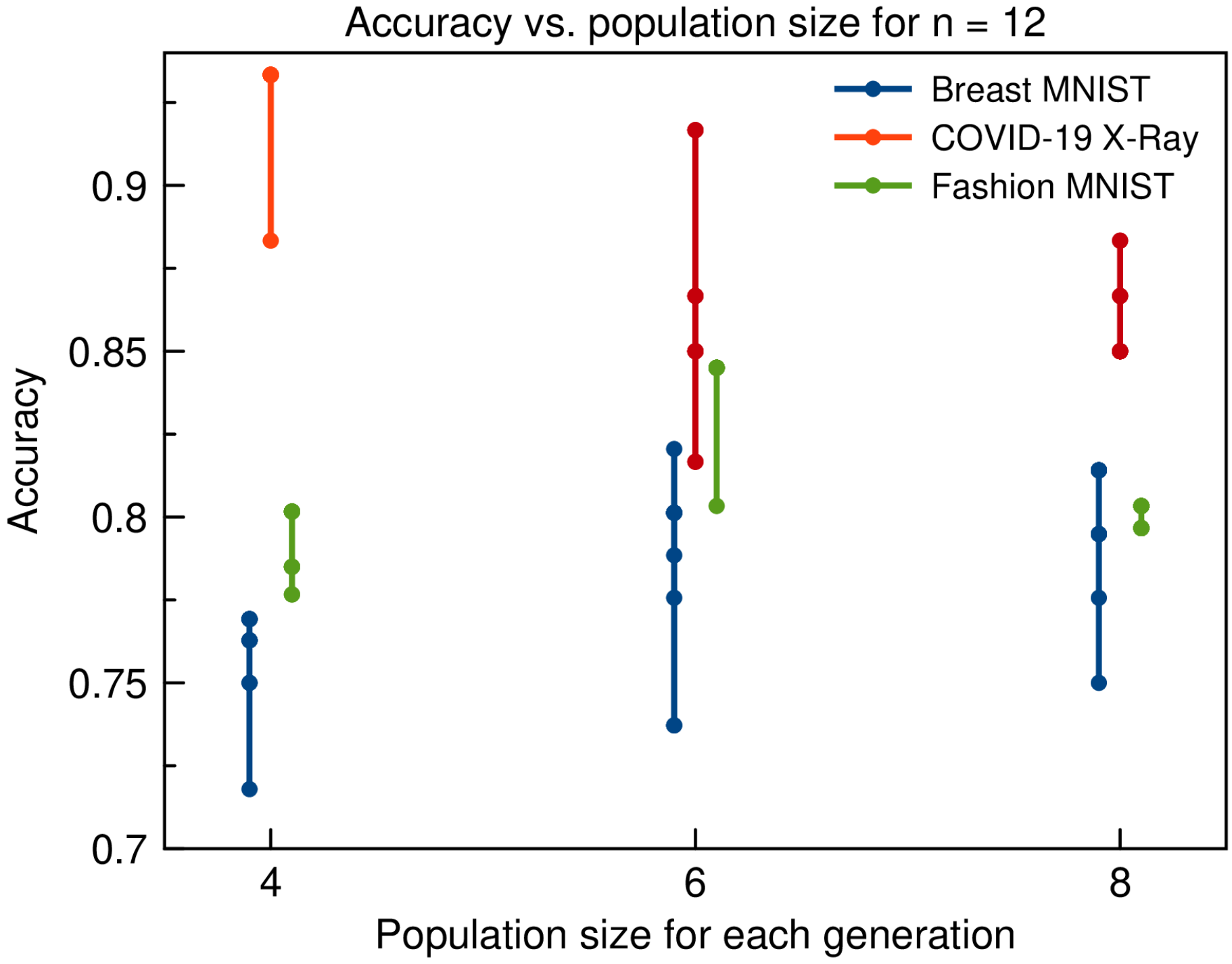

**Figure S9:** Accuracy values versus population size for a fixed number of generations of $n = 12$. Evolution of maximum accuracy through generations are given for population sizes of 4, 6, and 8.

Additionally, to examine how the accuracy changes under the condition that the number of random projection kernels evaluated during GA is kept constant at around 40, we performed experiments with $n = 5, p = 8$, and $n = 7, p = 6$. The evolution of maximum accuracy through generations is shown in Fig. S10 along with other experiments having different GA parameters. In the graph, while the line type (dashed or straight) differentiates between $n$ values, different $p$ values are color coded.

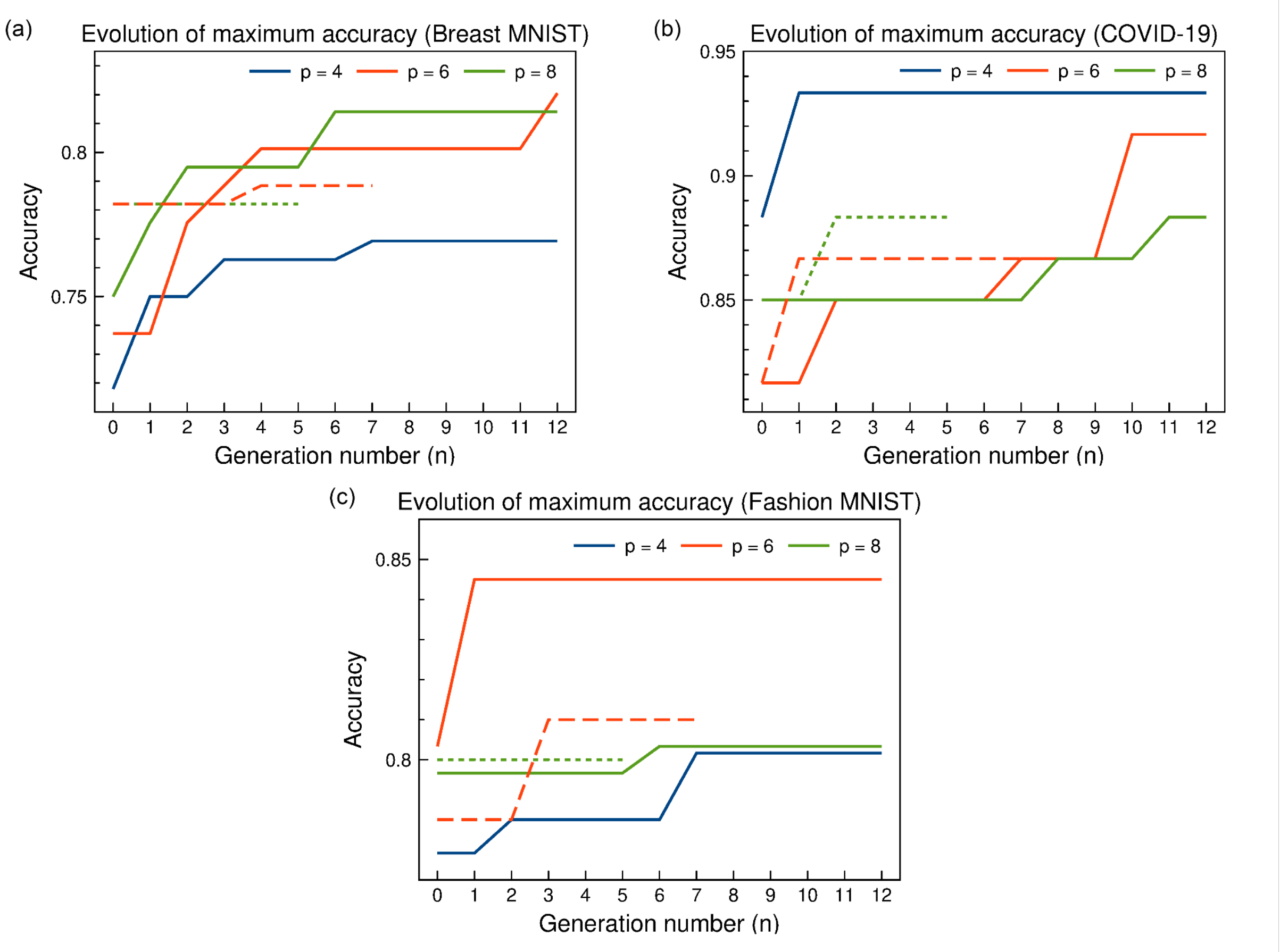

**Figure S10:** Evolution of maximum accuracy with different GA parameters for the (a) Breast MNIST, (b) COVID-19 X-Ray, and (c) Fashion MNIST datasets. $p$ and $n$ denote the population size and the number of generations, respectively.

## Supplementary Discussion 11: Detailed Visualization of Clustering Performance

Fig. S11 shows 3D LDA results on the randomly-mapped subset of the Fashion MNIST dataset for the best GA iteration from various elevation and azimuthal angles.

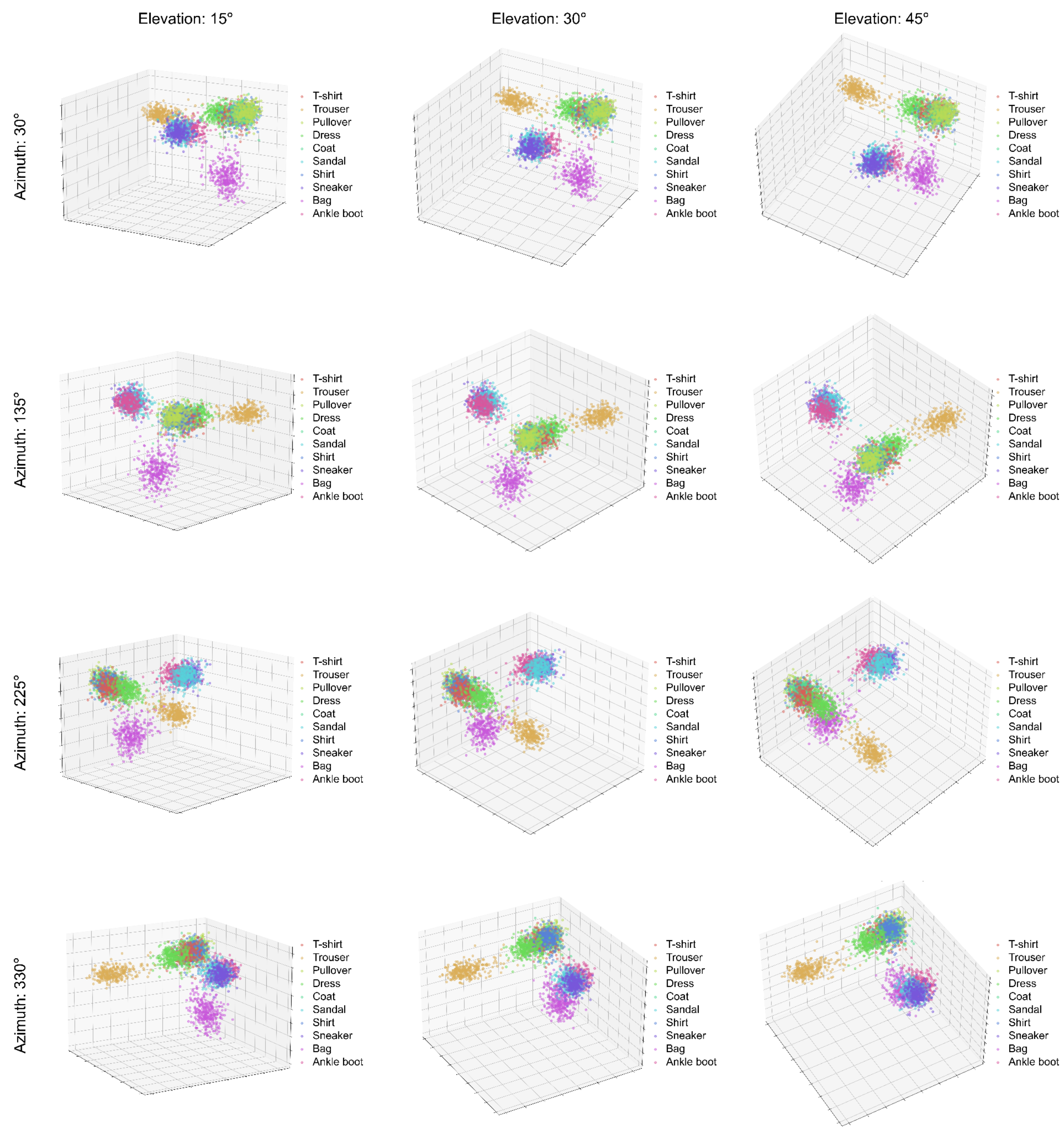

**Figure S11:** 3D linear discriminant analysis (LDA) results on the randomly-mapped Fashion MNIST dataset for the best genetic algorithm (GA) iteration for elevation angles of $15^\circ$, $30^\circ$ and $45^\circ$ and azimuthal angles of $30^\circ$, $135^\circ$, $225^\circ$ and $330^\circ$.

## Supplementary Discussion 12: Effects of Data Compression on Performance

When applying local averaging to randomly mapped data to reduce the number of pixels used for ridge classification, we fixed a pool size of $(20,16)$. Fig. S12 shows alternative pool sizes and the corresponding number of pixels along with ridge classification accuracies.

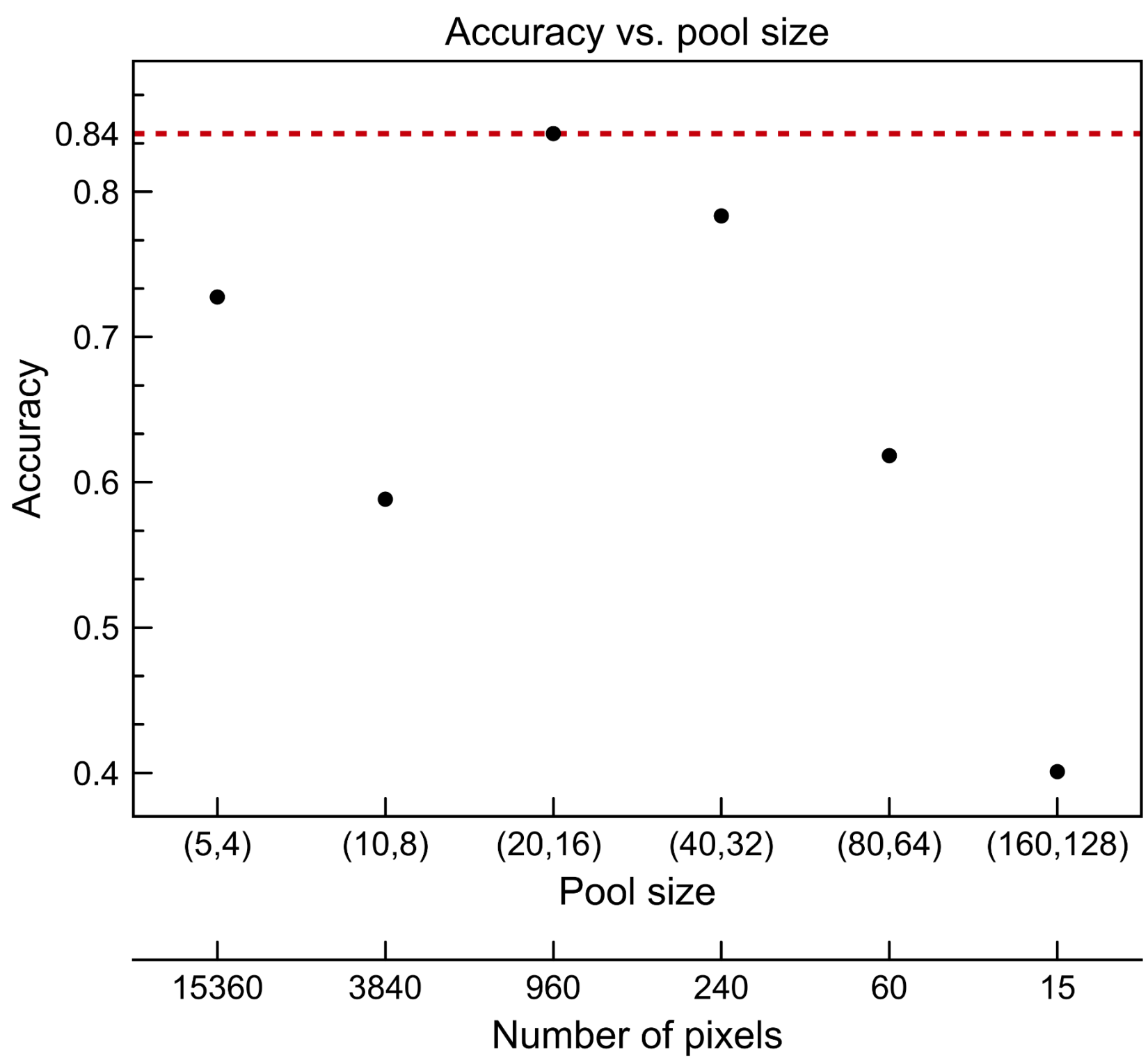

**Figure S12:** Ridge classification accuracies corresponding to different pool sizes used in local averaging for a randomly mapped subset of the Fashion MNIST dataset consisting of 6000 samples. The maximum accuracy is obtained for the aforementioned pool size of $(20,16)$.

Data compression can also be carried out through reducing the bit depth in captures obtained from randomly mapped inputs. In our experimental setup, we used a camera capable of recording 8-bit grayscale images. Although practically not possible because of the way digital computers process data, we emulated a camera capable of representing images using less bits to test our approach in data-scarce settings by mapping the range $0 - 255$ to a smaller range. If such an operation were possible, faster data rates would be enabled without sacrificing the classification accuracy. Fig. S13 shows GA results for different bit depths. We conclude that with suitable hardware tailored for modern machine vision tasks, our method can offer improvements in classification accuracy.

Although the positive effects of reducing the bit depth on the accuracy seem counterintuitive, we associated different levels of quantizing the resulting images with nonlinearities of different power. As is known from neural networks, linear operations alone are not sufficient for generalization. Therefore, reducing the bit depth improves classification accuracy, and this reduction can be interpreted as employing ReLU (or sigmoid) functions of different slopes. However, further investigation is needed to uncover the exact relation between different bit depth levels and conventional nonlinearities.

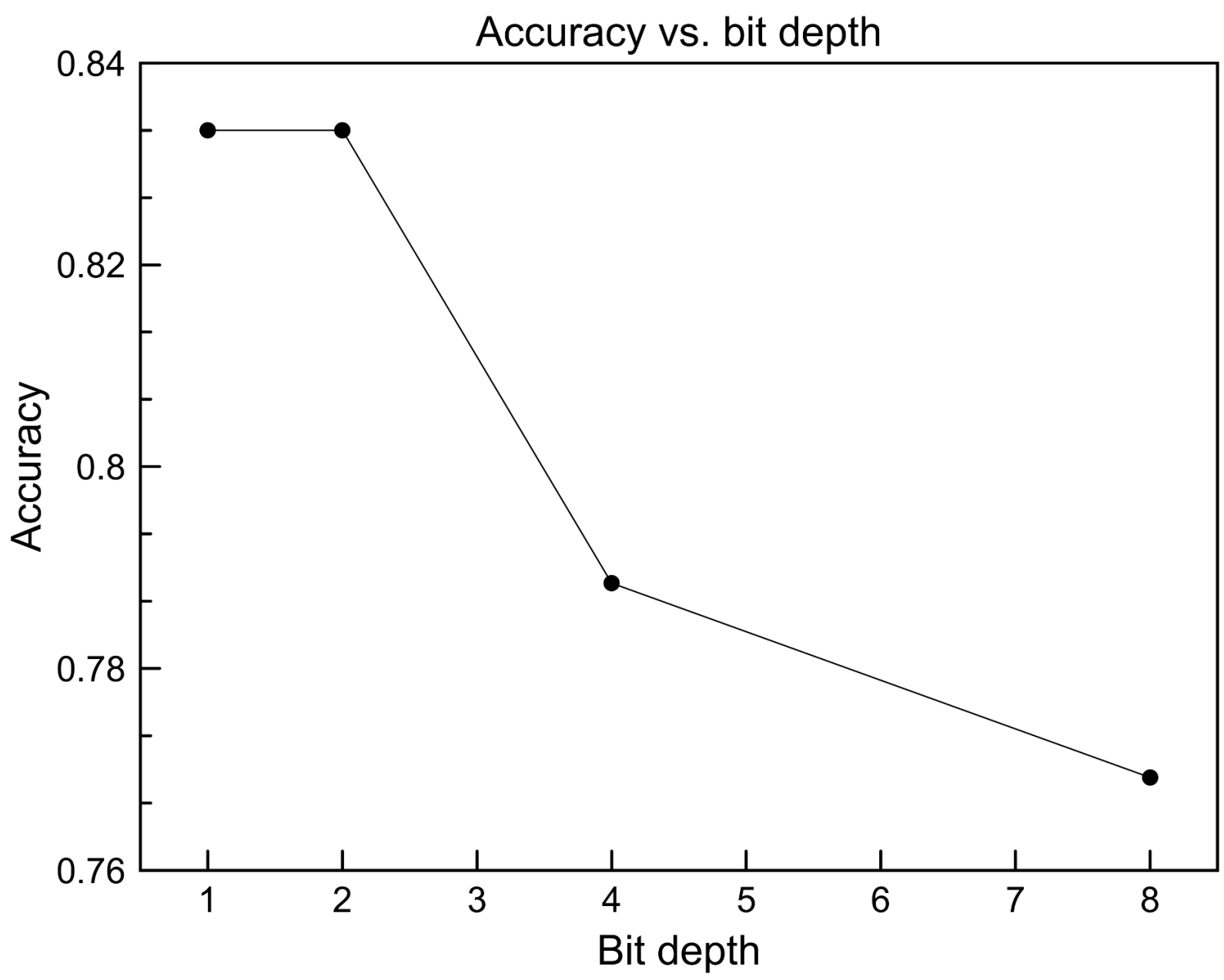

**Figure S13:** Dependence of ridge classification accuracy on the bit depth of captures in the Breast MNIST dataset. Separate experiments were carried out where 8-bit original captures were normalized and represented with 1, 2, and 4 bits. A maximum accuracy of 83.33% was obtained for bit depths of 1 and 2.

Table S1 demonstrates the effectiveness of our approach by comparing the classification accuracies obtained by (i) processing the randomly-mapped, 8-bit data and (ii) running GA from scratch (Fig. S13) to yield representations using less bits. These cases are denoted by the “accuracy without GA” and “accuracy with GA” rows, respectively. Without GA, after reducing 8-bit captures to 4-, 2-, and 1-bits and applying pooling and ridge classification, we observed small to no changes in classification accuracies, whereas when we let GA search for the optimal random mapping in different quantization schemes, we observed relatively higher improvements. This implies the adaptive nature of our approach to cases where high bitrates are desired when bit depth can be sacrificed. Modern machine vision tasks requiring real-time performance, therefore, could be a future application where we can test our programmable reservoir.

**Table S1:** Comparison of different quantization schemes in terms of classification accuracy with and without running genetic algorithm (GA) for the Breast MNIST dataset.

| | 8-bits | 4-bits | 2-bits | 1-bit |
|---|---|---|---|---|
| **Accuracy without GA** | 76.92% | 76.92% | 77.56% | 76.28% |
| **Accuracy with GA** | - | 78.85% | 83.33% | 83.33% |

**Supplementary Discussion 13: Benchmarking Accuracy Levels**

To provide a broader perspective, we compared our approach with other popular approaches on the same datasets, taking into account the number of parameters used in each approach. This comparison is presented in Table S2 below. From the table, one can conclude that programmable optical random neural networks generalize well across different datasets when the number of parameters are constrained.

**Table S2:** Comparison of different approaches in terms of their classification accuracy and number of parameters for different datasets.

| Dataset | Approach | Test accuracy | Number of parameters |
|---|---|---|---|
| Breast MNIST | Our approach | 82.05% | 961 |
| | ResNet-18 (3) | 86.3% | ~11 million |
| | Ridge regression | 66.67% | 784 |
| | PCA + Ridge | 72.44% | 2 |
| COVID-19 X-Ray | Our approach | 90.14% | 961 |
| | ResNet-101 (4) | 95.58% | ~44.5 million |
| | Ridge regression | 74.85% | 1000 |
| | PCA + Ridge | 60.97% | 2 |
| Fashion MNIST | Our approach | 83.06% | 961 |
| | ResNet-18 (5) | 94.9% | ~11 million |
| | Ridge regression | 81.13% | 784 |

| | PCA + Ridge | 39.68% | 2 |
|---|---|---|---|
| Retinal disease classification | Our approach | 87.70% | 961 |
| | Vanilla CNN | Not trainable | ~468.9 million |
| | Ridge regression | 85.65% | 960 |
| | PCA + Ridge | 91.73% | 2 |
| RS_C11 | Our approach | 21.05% | 961 |
| | Ridge regression | 8.91% | 1024 |
| | PCA + Ridge | 20.65% | 2 |
| AID | Our approach | 25.95% | 961 |
| | VGG-16 (7) | ~90% | ~138 million |
| | Ridge regression | 12.9% | 900 |
| | PCA + Ridge | 7.45% | 2 |

**Supplementary Discussion 14: Energy Consumption Across Computing Platforms**

We now provide quantitative metrics regarding the energy consumption in our optical computing platform and conventional GPU-based computing platforms. In our setup, energy consumption is only due to the light source, spatial light modulator, stepper motor, camera, and the CPU used to implement ridge classification, which have power ratings of 4.6 W, 50 W, 1.2 W, 1.1 W, and 25 W, respectively, totalling 81.9 W. Meanwhile, an NVIDIA RTX 4070 graphics card draws 200 W of power.